# Basal Slip in Laves Phases:
# the Synchroshear Dislocation

Julien Guénolé [a], Fatim-Zahra Mouhib [a], Liam Huber [b], Blazej Grabowski [b], Sandra Korte-Kerzel [a]

[a] Institute of Physical Metallurgy and Metal Physics, RWTH Aachen University,
D-52056 Aachen, Germany
[b] Max-Planck-Institut für Eisenforschung GmbH, Max-Planck-Str. 1, D-40237 Düsseldorf, Germany

Two different mechanisms have been reported in previous *ab initio* studies to describe basal slip in complex intermetallic Laves phases: synchroshear and undulating slip. To date, no clear answer has been given on which is the energetically favourable mechanism and whether either of them could effectively propagate as a dislocation. Using classical atomistic simulations supported by *ab initio* calculations, the present work removes the ambiguity and shows that the two mechanisms are, in fact, identical. Furthermore, we establish synchroshear as the mechanism for propagating dislocations within the basal plane in Laves phases.

Laves phases are intermetallic compounds that form in many alloys and can play a crucial role in their deformation [1,2]. Understanding the *intrinsic* deformation behaviour of Laves phases is critical and many activities have been initiated in that direction [3–12]. Laves phases can be cubic (C15) or hexagonal (C14 and C36) layered structures with the archetypal chemical compositions of $Cu_2Mg$, $Zn_2Mg$ and $Ni_2Mg$. The lattice consists of two structural units: a single layer formed of large atoms and a triple layer mixing small and large atoms. For the hexagonal structures, these layers form the basal planes. In their pioneering work, Hazzledine *et al.* [3,4] predicted basal-slip to occur *via* a synchroshear mechanism [13] that reorganises the stacking of the layered structure by the motion of a synchroshockley dislocation within the triple layer, that is the synchronous glide of two ordinary Shockleys on adjacent parallel atomic planes. The Burgers vector of a synchroshockley dislocation is $\boldsymbol{b_s} = \frac{1}{3}\langle 10\bar{1}0\rangle$ and the glide of two synchroshockleys leads to the regular Burgers vector $\boldsymbol{b} = \frac{1}{3}\langle 11\bar{2}0\rangle$ [3]. More details on the Laves phase crystallography can be found in the references or in the Supplementary material.

The first experimental observation of a basal dislocation core in a Laves phase was reported by Chisholm *et al.* [5] in C14 $Cr_2Hf$ and was attributed to a synchroshockley. Atomistic studies of the basal slip in C14 $Cr_2Nb$ have revealed two apparently different slip mechanisms. Based on *ab initio* calculations, Vedmedenko *et al.* [7] reported the experimentally observed synchroshear mechanism (SS), whereas Zhang *et al.* [8] reported a new mechanism called undulating slip (US). SS is described as the synchronous glide of two partials in the basal plane without any out-of-plane motion. Although no details on the dynamics of this mechanism are known, its activation energy is significantly lower than for classical crystallographic shear [7]. US also consists of the glide of two partials in the basal plane, but is assumed to have a different Burgers vector than SS and proceeds dynamically in three stages: 1) an elastic motion (termed crystallographic slip in [8]), 2) a plastic motion involving breaking of atomic bonds and atomic shuffling, and again 3) elastic motion/crystallographic slip. Note that the term "crystallographic slip" used in [8] can be seen as controversial since there is no atomic sliding at this stage. So far, no study has attempted to differentiate SS and US in any detail. In the present work, we address this point and show that US is in fact not a new mechanism, but rather the same as SS. This is accomplished using the nudge elastic band (NEB) approach and classical atomistic simulations supported by *ab initio* calculations.

The NEB method enables the calculation of both the minimum energy path (MEP) and the associated activation energy (or energy barrier $E_a$) of a mechanism [14–16]. NEB makes no assumption about the MEP besides the initial guess for the path, for which we use a linear interpolation of particle positions between an initial and a finale stable configuration. It is thus a well suited tool to assess unknown dislocation mechanisms, as for nucleation [17] or core transformations [18]. Relying on a recently developed interatomic potential [19], we study here basal slip in the Laves phase. While NEB has been used previously to study this effect with *ab initio* methods [7,8], these studies were limited to relatively small system sizes and the partial slip only, i.e. the transformation of a triple layer from a C14 to a C15 stacking. This restricts the degrees of freedom available for the system to use during slip. By using a classical potential, we can investigate larger system sizes and avoid imposing such constraints. The potential uses the modified embedded energy method (MEAM) framework to accurately describe the mechanical properties of Mg-Ca compounds, in particular the C14 $Mg_2Ca$ Laves phase [19]. Additionally, we have validated bulk properties of this phase against *ab initio* calculations performed with the Vienna Ab-initio Simulation Package (VASP) using the projector augmented wave method (See Supplementary materials).

Figure 1a shows the generalized stacking fault energy (GSFE) for the basal slip as calculated in this work with the MEAM potential. Slip directions ⟨11$\bar{2}$0⟩ and ⟨10$\bar{1}$0⟩ are indicated by red and blue solid lines, respectively. The red and blue lines also correspond to the linearly interpolated initial configurations for the NEB in the case of slip in ⟨11$\bar{2}$0⟩ and ⟨10$\bar{1}$0⟩ directions, respectively. Note however that, while the GSFE contains the energy path of the crystallographic slip mechanism as considered in the NEB calculations, the calculation procedures are different. The GSFE is computed with the commonly used *free* boundary conditions in the direction normal to the glide plane, whereas the NEB calculations employ *periodic* boundary conditions in all directions. To ensure a constant strain state throughout the NEB calculations, the initial configuration has been elastically strained with a shear equivalent to a full Burgers vector such as to correspond to the plastic strain of the final configuration. Thus, configuration A exhibits a higher energy than configuration B in all computed NEB energy profiles (Figures 1b, 2a, 3).

In order to assess the energy associated with mechanisms other than pure crystallographic slip, we performed NEB minimizations to reduce the force norms below 0.02 eV/Å using the quickmin algorithm as implemented in Lammps [16,20]. Due to the complexity of the system, it was necessary to add a spring force perpendicular to the path to converge the calculations, as proposed by Maras *et al.* [16]. The initial NEB configurations have been linearly interpolated between the two perfect C14 $Mg_2Ca$ configurations labelled A and B (Figure 1a, red line), consisting of a fully periodic system of 1×2×20 unit cells (480 atoms) with the z-axis normal to the basal plane. The computed MEP (Figure 1b, blue line) exhibits a profile with two maxima and an energy barrier lower than for the crystallographic slip. The associated atomistic mechanism within the triple layer is shown in Figure 1d. The Ca atoms slip in $\langle 10\bar{1}0 \rangle$ directions for each of the MEP peaks (configurations 1 and 3), corresponding to a motion of the Mg atoms along $\langle 1\bar{1}00 \rangle$ directions. For each peak, the *absolute* glide of the Ca atoms corresponds to $\frac{1}{6}\langle 10\bar{1}0 \rangle$, but the glide of the top-layer Ca atoms *relative* to the bottom-layer Ca atoms corresponds exactly to $\frac{1}{3}\langle 10\bar{1}0 \rangle$. The total shear is $\frac{1}{3}\langle 11\bar{2}0 \rangle$ as expected. Simultaneously, the glide of the middle-layer Mg atoms is $\frac{1}{3}\langle 10\bar{1}0 \rangle$. Without any assumption on the mechanism beside the Burgers vector of the full slip and that the linear interpolation lies in the same energy well as the MEP, it is clear that the MEP of a full shear in the C14 $Mg_2Al$ structure corresponds exactly to two successive SSs [5,7].

Based on the above results and a re-interpretation of results from Ref. [8], we can now identify US as the same mechanism as SS. First, we focus on the energy barriers, $E_b$. The $E_b$ computed in Ref. [8] for US in C14 $Cr_2Nb$ (0.28 eV/$Cr_2Nb$) is nearly the same as the $E_b$ of SS in the same system (0.29 eV/$Cr_2Nb$) [7]. Further, the $E_b$ calculated for SS in Ref. [8] (0.41 eV/$Cr_2Nb$) is surprisingly close to the peak energy of the *unrelaxed* SS (0.42 eV/$Cr_2Nb$) [7]. This indicates that the MEP attributed in Ref. [8] to US is in fact the one of *relaxed* SS. Now we focus on the Burgers vector. For US, the plastic motion of the intermediate atomic layer was observed to be *perpendicular* to the slip direction, i.e., along $\langle 11\bar{2}0 \rangle$ [8]. This motion thus seems to be incompatible with that of SS, which is purely along $\langle 10\bar{1}0 \rangle$. However, the plastic motion for the US is preceded and followed by a purely elastic motion. To evaluate the Burgers vector of the slip, the full process including the elastic and plastic motion needs to be considered. Hence, summing up the elastic motion with the plastic motion, the total motion of the intermediate layer is along the $\langle 10\bar{1}0 \rangle$ direction, exactly as predicted for SS. We thus conclude that all features of US are already captured by the dynamics of SS and therefore US is not a new mechanism for basal slip in C14 structures.

To investigate the SS dynamics further, we studied slip in the $\langle 10\bar{1}0 \rangle$ direction in greater detail. Figure 2a shows the energy path of the crystallographic slip as well as the MEP of the SS mechanism in the $\langle 10\bar{1}0 \rangle$ direction. The SS MEP has been obtained by NEB, with the initial configurations linearly interpolated between perfect and faulted C14 $Mg_2Ca$ configurations A and B, respectively. By faulted, we mean that one of the triple layers of the C14 structure is in the C15 configuration, i.e. it has been displaced according to the SS mechanism. As in our previous simulations, the initial configuration A has been strained and the periodic simulation box consists of 1×2×20 units cells, i.e., two $Mg_2Ca$ units in the basal plane (Figure 2b,c dashed box).

The $MEP^{48}$ and $MEP^{177}$ obtained with 48 and 177 intermediate configurations, respectively, exhibit completely different profiles (Figure 2a). Using additional intermediate configurations enables the NEB minimization to find a lower energy transition mechanism. Thus the $MEP^{177}$ gives the most energetically favourable mechanism. The $MEP^{48}$ and $MEP^{177}$ consist of two and three maxima, respectively. This difference originates from the atomistic mechanisms detailed in Figure 2b,c. The $MEP^{48}$ exhibits a broad peak that corresponds to the atomic shuffling of the Mg and Ca atoms, preceded and followed by purely elastic shear (Figure 2b). Meanwhile, the first maximum in the $MEP^{177}$ corresponds to the shuffling of the Mg/Ca atoms in the first $Mg_2Ca$ unit (Figure 2c, configuration 1). The second maximum is related to a purely elastic shear of the entire system that occurs between configuration 2 and 3. The third maximum is finally related to the shuffling of the Mg/Ca atoms in the second $Mg_2Ca$ unit (Figure 2c, configuration 5). In both cases, the overall mechanism is a synchroshear.

In parallel, we have performed *ab initio* NEB calculations in a C14 $Mg_2Ca$ system of 1×2×2 unit cells (48 atoms). While significantly more computationally expensive, this framework enables more accurate calculations than semi-empirical potentials. We observe that the MEP of the basal slip remains associated with SS, and that the associated energy barrier is in good agreement with the MEAM potential (see Section 3.3 and Movie 1 in Supplementary materials). These qualitative and quantitative DFT calculations validate the results we obtained with the MEAM potential.

It appears that SS as computed by NEB is a slip that is divided into two phases: an atomic shuffling and an elastic straining. With sufficient degrees of freedom (DoF), a lower energy path than collective SS can be found, by inducing an asynchronous atomic shuffling of the different C14 units within the basal plane. We confirmed this last statement by studying larger systems, which intrinsically increase the DoF of the NEB process. Figure 3 shows the MEP for a full-slip $\frac{1}{3}\langle 11\bar{2}0 \rangle$ in the basal plane of systems with 1×2×20 (480 atoms), 2×4×20 (1920 atoms) and 4×8×20 (7680 atoms) unit cells, resulting in 2, 8 and 32 Mg$_2$Ca units in the basal plane, respectively. The smallest system is identical to the one presented above (Figure 1b). Note that varying the number of NEB intermediate configurations increases the number of local maxima in the MEP. This is related to the shuffling of individual Mg$_2$Ca units, but does not alter the mechanism and does not change the value of $E_b$ significantly (Figure 3, and Supplementary materials). The $E_b$ clearly decreases with the number of Mg$_2$Ca units in the basal plane, consistent with the asynchronous process described above. The overall associated mechanism is always a double SS. The decrease of $E_b$ originates in the details of the slip, in particular its irregularity (see Movies 2 and 3 in Supplementary materials). Indeed, the motion of a particular atom is required to initiate the slip but is unfavourable since it increases the overall energy of the system. It is then more efficient to initiate the motion on one part of the system, and then to propagate it through the entire plane. This is the fundamental reason for dislocations as plasticity carriers in crystalline materials [21] and therefore confirms SS as a robust mechanism for dislocation propagation in bulk C14 Mg$_2$Ca.

To validate SS as the most probable mechanism for dislocation propagation, we have performed NEB calculations of the slip in a compressed nano-pillar (Figure 4). The considered system was a cylinder with diameters of 8 nm (40,470 atoms) cut out of a C14 Mg$_2$Ca bulk phase. The pillar axis was oriented along $\langle\bar{1}2\bar{1}\bar{3}\rangle$ and strained by applying force fields at each end. The initial configuration was a non-faulted pillar, strained with a magnitude equivalent to a full slip, corresponding to an axial stress of approximately 250 MPa. Note that the intrinsic local stress was homogenous throughout the pillar (see Supplementary materials). The final configuration was a nearly stress-free pillar that has undergone a full slip of $\frac{1}{3}\langle11\bar{2}0\rangle$ in the basal plane under maximum Schmid factor (Figure 4b). Our NEB calculations clearly show the nucleation of slip from the surface, followed by its propagation via SS (Figure 4c and Movie 4 in Supplementary materials). In particular, the first (Figure 4c-1) and second (Figure 4c-2) synchroshockley dislocations propagate by both the typical atomic displacements of the SS mechanism described above and the atomic von Mises shear strain [22,23]. The associated MEP exhibits a characteristic two-maxima profile corresponding to two successive SS (Figure 3a). The details of the dislocation propagation remain unknown and suggested mechanisms, such as the vacancy-aided kink propagation [24], need to be further investigated. Nonetheless, with this NEB approach on a nano-pillar, we demonstrated the motion of a basal dislocation by SS.

The confirmation of the synchroshear mechanism in the present work as the most favourable mechanism for basal slip in a Laves phase is an important insight for future experimental and theoretical investigations of plasticity in general complex intermetallics. The final validation of SS will require a direct comparison between nano-mechanical experiments and nano-scale simulations at the experimental temperature.

**Acknowledgements**

Simulations were performed with computing resources granted by RWTH Aachen University under projects rwth0297 and thes0371. This project has received funding from the European Research Council (ERC) under the European Union's Horizon 2020 research and innovation programme (grant agreement No 639211).

[1] T.M. Pollock, Science 328 (2010) 986–7.
[2] A.J. Knowles, A. Bhowmik, S. Purkayastha, N.G. Jones, F. Giuliani, W.J. Clegg, D. Dye, H.J. Stone, Scr. Mater. 140 (2017) 59–62.
[3] P.M. Hazzledine, K.S. Kumar, D.B. Miracle, A.G. Jackson, MRS Proc. 288 (1992) 591.
[4] P.M. Hazzledine, P. Pirouz, Scr. Metall. Mater. 28 (1993) 1277–1282.
[5] M.F. Chisholm, S. Kumar, P. Hazzledine, Science (80-. ). 307 (2005) 701–703.
[6] Y. Zhong, J. Liu, R.A. Witt, Y. ho Sohn, Z.K. Liu, Scr. Mater. 55 (2006) 573–576.
[7] O. Vedmedenko, F. Rösch, C. Elsässer, Acta Mater. 56 (2008) 4984–4992.
[8] W. Zhang, R. Yu, K. Du, Z. Cheng, J. Zhu, H. Ye, Phys. Rev. Lett. 106 (2011) 2–5.
[9] Z.Q. Yang, M.F. Chisholm, B. Yang, X.L. Ma, Y.J. Wang, M.J. Zhuo, S.J. Pennycook, Acta Mater. 60 (2012) 2637–2646.
[10] N. Takata, H. Ghassemi Armaki, Y. Terada, M. Takeyama, K.S. Kumar, Scr. Mater. 68 (2013) 615–618.
[11] Y. Liu, W.C. Hu, D.J. Li, K. Li, H.L. Jin, Y.X. Xu, C.S. Xu, X.Q. Zeng, Comput. Mater. Sci. 97 (2015) 75–85.
[12] L. Liu, P. Shen, X. Wu, R. Wang, W. Li, Q. Liu, Comput. Mater. Sci. 140 (2017) 334–343.
[13] M.L. Kronberg, Acta Metall. 5 (1957) 507–524.
[14] G. Henkelman, B.P. Uberuaga, H. Jónsson, J. Chem. Phys. 113 (2000) 9901.
[15] G. Henkelman, H. Jónsson, J. Chem. Phys. 113 (2000) 9978.
[16] E. Maras, O. Trushin, A. Stukowski, T. Ala-Nissila, H. Jónsson, Comput. Phys. Commun. 205 (2016) 13–21.
[17] S. Brochard, P. Hirel, L. Pizzagalli, J. Godet, Acta Mater. 58 (2010) 4182–4190.
[18] J. Guénolé, J. Godet, L. Pizzagalli, Model. Simul. Mater. Sci. Eng. 18 (2010).
[19] K.-H. Kim, J.B. Jeon, B.-J. Lee, Calphad 48 (2015) 27–34.
[20] S. Plimpton, J. Comput. Phys. 117 (1995) 1–19.
[21] P.M. Anderson, J.P. Hirth, J. Lothe, Theory of Dislocations, Cambridge University Press, 2017.
[22] M.L. Falk, J.S. Langer, Phys. Rev. E 57 (1998) 7192–7205.
[23] F. Shimizu, S. Ogata, J. Li, Mater. Trans. 48 (2007) 2923–2927.
[24] K.S. Kumar, P.M. Hazzledine, Intermetallics 12 (2004) 763–770.

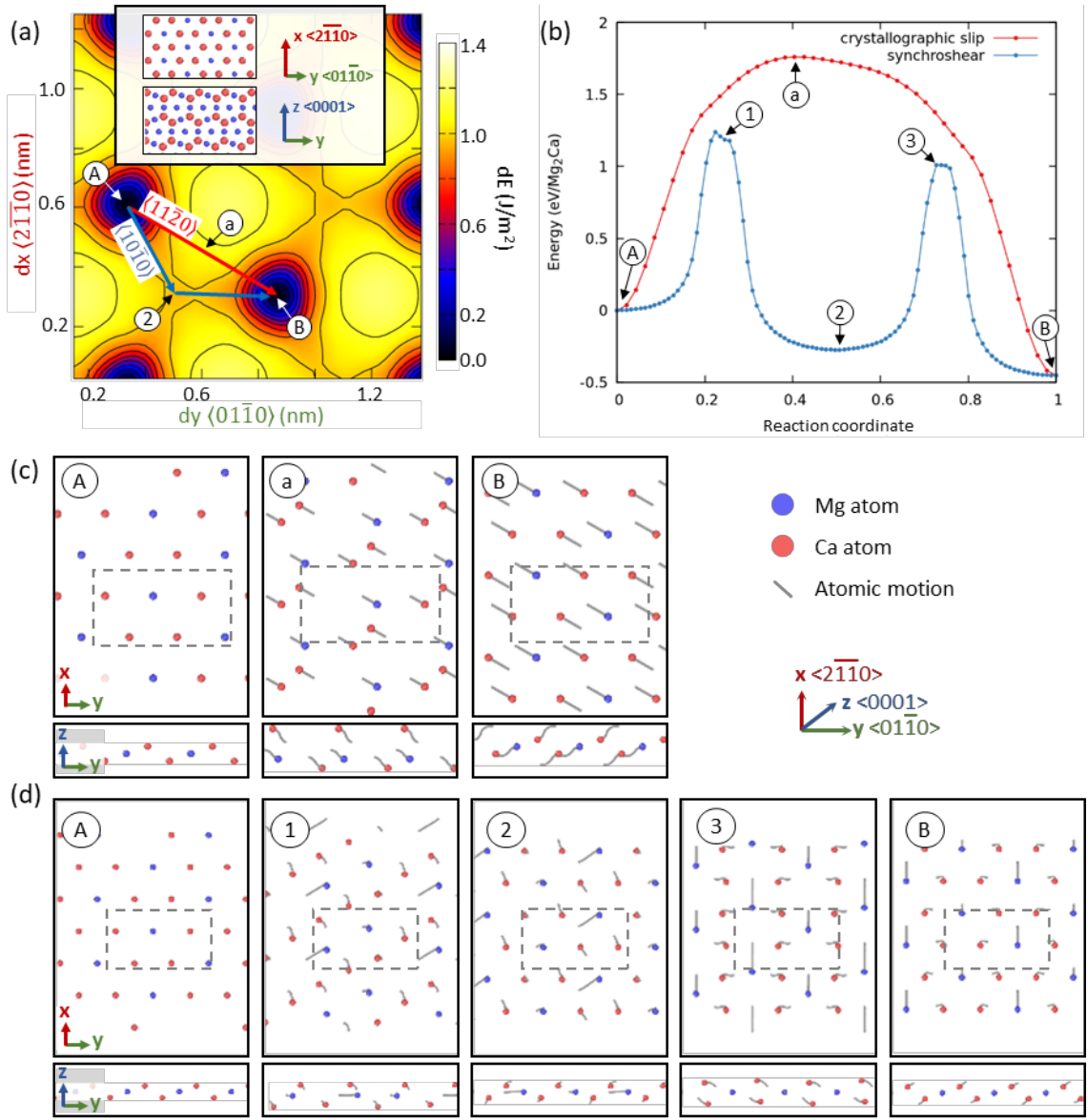

Figure 1: Basal slip in the C14 Mg$_2$Ca Laves phase. (a) Generalized stacking fault energy (GSFE) surface obtain with the MEAM potential. Slip directions ⟨11$\bar{2}$0⟩ and ⟨10$\bar{1}$0⟩ are indicated with red and blue arrows, respectively. Inset: detail of the layered structure. (b) Energy path for the full slip with two mechanisms: crystallographic slip (red curve) and synchroshear (blue curve). The synchroshear is the minimum energy path. The overall change from reaction coordinate 0 to 1 corresponds to the full slip $\frac{1}{3}$⟨11$\bar{2}$0⟩. Energy in eV per Mg$_2$Ca unit. Mg$_2$Ca units divide the basal plane in areas formed of 2 Mg and 1 Ca, each belonging in a different sublayer of the triple layer. Snapshots of the corresponding mechanisms for (c) the crystallographic slip and (d) the synchroshear. The upper (lower) boxes show the perpendicular (parallel) views of the basal plane. The boundary of the periodic simulation box is indicated as a dashed box. The motion of the atoms from the previous snapshot are indicated with grey lines. Mg and Ca atoms are coloured in blue and red, respectively.

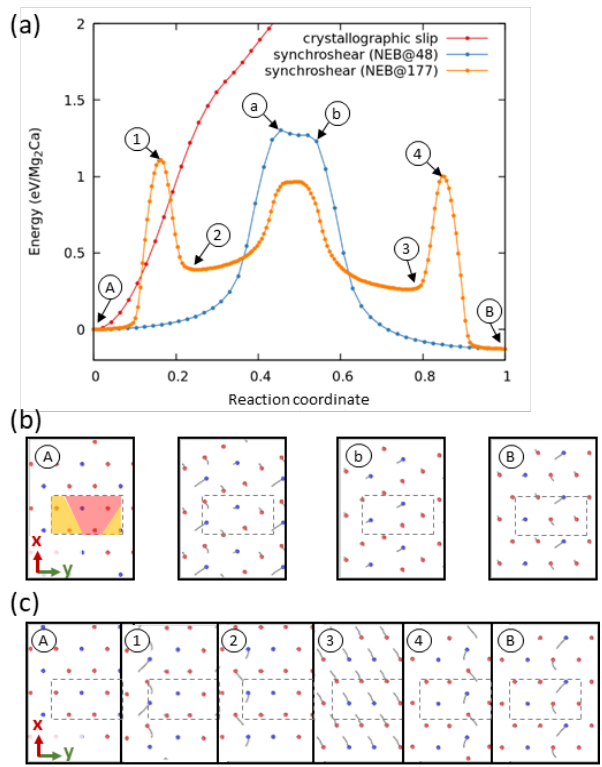

Figure 2: Propagating basal slip in the C14 Mg$_2$Ca Laves phase. (a) Energy path for the partial slip with two mechanisms: crystallographic slip (red curve) and single synchroshear (red and orange curves). The synchroshear is the minimum energy path, as obtained by NEB with 48 and 177 intermediate images for the blue and orange curve, respectively. The overall change from reaction coordinate 0 to 1 corresponds to the partial slip $\frac{1}{3}\langle 10\bar{1}0\rangle$. Basal plane snapshots of the corresponding synchroshear mechanisms with (b) 48 and (c) 177 NEB intermediate configurations. The inset numbers in black circles correspond to the intermediate configurations in (a). The periodic simulation box consisting on two Mg$_2$Ca units is indicated as a dashed box. Units' areas are highlighted in yellow and red transparent filling color. The motion of the atoms from the previous snapshot are indicated with grey lines. The crystallographic orientations are the same as in Figure 1.

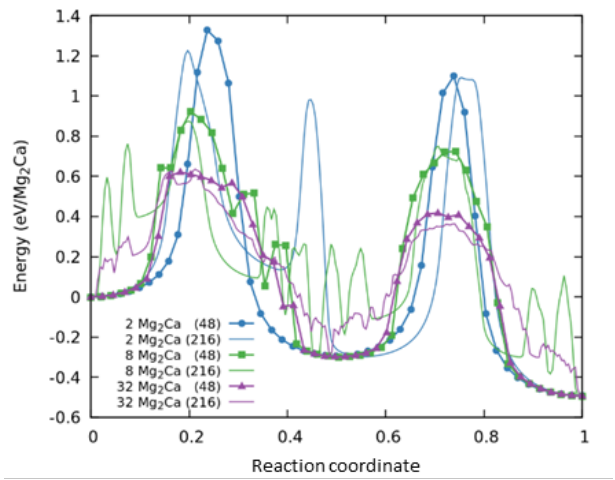

Figure 3: Basal slip in the C14 Mg$_2$Ca Laves phase in bulk systems of different periodic dimension. Minimum energy path computed for a full slip by NEB with 48 (216) images in thick (thin) solid lines. The overall change from reaction coordinate 0 to 1 corresponds to the full slip $\frac{1}{3}\langle 11\bar{2}0\rangle$. Note that the blue thick curve is identical to that in Figure 1b.

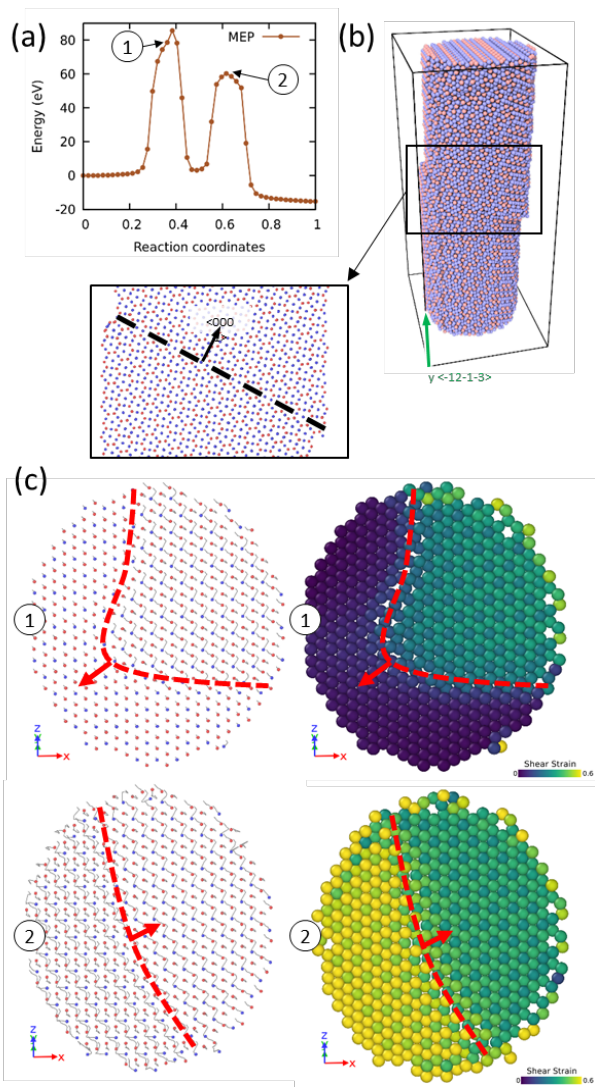

Figure 4: Basal dislocation in C14 Mg$_2$Ca Laves phase nano-pillars. (a) Activation energy and minimum energy path for the nucleation of a dislocation in the basal plane. (b) Snapshot of the 8 nm nano-pillar after full slip. Inset of the glide plane and surface steps. (c) Atomic displacements and atomic shear strain of the triple-layer atoms involved in the slip mechanism, for two intermediate configurations indicated in (a). Dislocation line and dislocation motion indicated with dashed red line and red arrow, respectively.

**Graphical abstract**

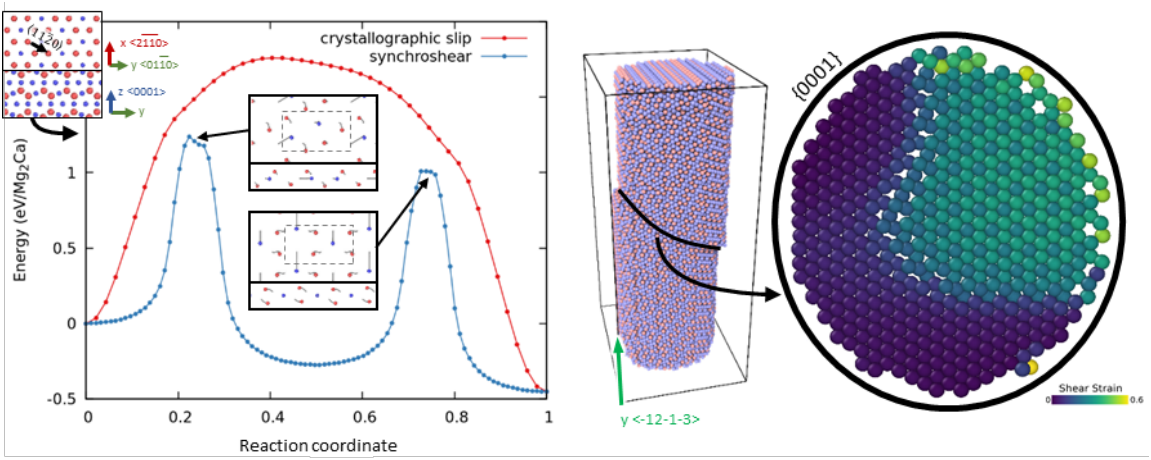